\documentclass[9pt,twocolumn,twoside]{opticajnl}
\journal{opticajournal} % use for journal or Optica Open submissions

\setboolean{shortarticle}{true}
\usepackage{lineno}
\title{Architecture-agnostic analysis of partially coherent light with programmable photonics}
\author[1]{Kevin Zelaya}
\author[1]{Matthew Markowitz}
\author[1]{Jonathan Friedman}
\author[1,*]{Mohammad-Ali Miri}

\affil[1]{Department of Electrical and Microelectronic Engineering, Rochester Institute of Technology, Rochester, New York 14623, USA}

\affil[*]{ali.miri@rit.edu}

\begin{abstract}
The precise characterization of the spatial degree of coherence of a radiation field is important for assessing its suitability for specific applications in optical communications, advanced imaging, and quantum information processing. However, measuring the full coherence matrix traditionally requires complex, phase-sensitive interferometric setups that are highly susceptible to noise and difficult to scale on integrated platforms. To address this, we propose an architecture-agnostic approach for analyzing partially coherent light that is compatible with any universal programmable photonic unitary circuit, regardless of its internal topology. Leveraging the Schur-Horn theorem, our method diagonalizes the output coherence matrix, enabling direct extraction of its eigenvalues from output power measurements alone. We numerically validate this framework across various universal topologies and demonstrate its efficacy even in under-parameterized, non-universal architectures with only minor loss in precision. Finally, our black-box optimization approach proves inherently resilient to arbitrary optical losses and component deviations, paving the way for robust, lower-depth, and programmable spatial coherence analyzers.
\end{abstract}

\setboolean{displaycopyright}{false} % Do not include copyright or licensing information in submission.

\begin{document}

\maketitle

% \section{Introduction}

\textit{Introduction} -- Optical coherence is a physical quantity that characterizes the phase correlation of a light wavefront. Two types of optical coherence are of interest: temporal and spatial coherence. Temporal coherence measures the stability of this phase along the direction of propagation over a specific period. In turn, spatial coherence describes the phase correlation between different points across the transverse profile of the wave, which ultimately determines the ability to form stable interference patterns. Indeed, the higher the coherence of the radiation field, the higher the contrast of interference patterns in diffraction experiments~\cite{mandel1996optical}, such as Young's double-slit and Michelson's experiments. In contrast, some applications require partially coherent fields to mitigate the scattered speckle patterns~\cite{lee2020light}. Among the different mechanisms for characterizing the spatial coherence of classical radiation fields, Photonic Integrated Circuits (PICs) are a particularly useful platform, as they enable the integration of complex optical components in a small form factor. This reduces the calibration effects that typically hinder characterization in free-space optics and also enables tunability and on-chip control of light, which is useful for post-fabrication calibration if manufacturing defects are present. 

Linear discrete unitary transformations form the basis of optical system analysis, as such transformations naturally result from light propagation in free space and dielectric media. Since general unitary matrices $U(N)$ can be factored into sequences of smaller-dimensional $U(2)$ transforms~\cite{reck1994experimental,de2018simple}, their implementation becomes feasible with optical components. Indeed, such a modular approach was proposed in free-space by Reck \textit{et al.} by using beam splitters, mirrors, and waveplates~\cite{reck1994experimental}. In integrated photonics, tunable $U(2)$ transformations can be implemented using directional couplers~\cite{miller2013self} and ring resonators~\cite{yi2021multi} as passive components, combined with phase shifters as active components. Various integrated topologies based on $U(2)$ decompositions have been explored in the literature; notable examples include the rectangular network by Clements \textit{et al.}~\cite{clements2016optimal} and diamond-like topologies~\cite{Shokraneh2020,rahbardar2023addressing}, which enable alternative strategies for integrating universal unitary transforms. To bypass the $U(2)$ decomposition, other successful approaches have utilized $N$-port couplers as passive components, interlacing them with phase shifters~\cite{tanomura2022scalable,zelaya2024goldilocks}. These designs can be deployed through waveguide arrays~\cite{friedman2025programmable} and multimode interferometers (MMIs)~\cite{pastor2021arbitrary}, enabling overparameterized unitaries and error-resilient performance~\cite{Markowitz2023auto}.

In this letter, we introduce an in-situ method for analyzing the spatial optical coherence of radiation fields using PICs, regardless of the internal topology. This enables an architecture-agnostic approach that treats the PIC as a black box and operates in nonideal scenarios. We explore this by numerically demonstrating that non-universal, low-depth, and lossy architectures can achieve coherence analysis with only a minor precision penalty, paving the way for faster, programmable solutions with a reduced physical footprint.

% \section{Results}
% \subsection{Theory}
\textit{Results} -- Throughout the letter, we consider a radiation field $\mathbf{E}(\mathbf{r},t)$ sampled at the spatial points $\mathbf{r}_{i}$, for $i\in\{1,\ldots,N\}$. The sampled points are encoded into a complex-valued vector $\mathbf{x}(t)\in\mathbb{C}^{N}$, with its $i$-th component containing, without loss of generality, a specific polarization component of the transverse electric field component at the point $\mathbf{r}_{i}$. In this form, $\mathbf{x}$ carries information about the spatial coherence of the original field, which is more conveniently represented through the \textit{coherence matrix}~\cite{born2013principles} defined as $\rho=\langle \mathbf{x}(t)\mathbf{x}(t)^{\dagger} \rangle$. Here, $\mathbf{x}^{\dagger}$ is the conjugate transpose of $\mathbf{x}$, and $\langle \cdot \rangle$ denotes the ensemble or time average of the radiation field. Thus, $\rho$ encodes spatial correlation between the different spatial sampling points. Mathematically, by assuming that fields are statistically stationary~\cite{mandel1996optical}, it is straightforward to see that the coherence matrix is both Hermitian and positive semi-definite. These two properties allow factoring the coherence matrix via its eigen-decomposition $\rho = V\Lambda V^{\dagger}$, with $V$ a unitary matrix ($V\in \mathcal{U}(N)$), and $\Lambda=\textnormal{diag}(\lambda_{1},\ldots,\lambda_{N})$ a diagonal matrix containing the positive semi-definite eigenvalues ($\lambda_{n}\geq 0$) sorted in non-increasing order ($\lambda_{n}\geq\lambda_{n+1}$, for $n\in\{1,\ldots,N-1\}$). 

%\textcolor{red}{The coherence matrix we work with is composed of ONLY second-order correlation components across different sampling points. Why are second-order correlations relevant, and what properties do they reveal? Applications?}

The coherence matrix we work with consists solely of second-order correlation components across different sampling points. These second-order correlations are fundamental because they directly quantify phase stability and mutual intensity across spatial locations, revealing the underlying spatial coherence of the field and its capacity to form stable interference patterns. Precise analysis of this spatial degree of coherence is particularly required in advanced imaging applications, such as mitigating crosstalk in optical coherence tomography~\cite{tomlins2005theory} and speckle patterns~\cite{lee2020light}.

% -------------------------------------------
\begin{figure*}[ht!]
    \centering
    \includegraphics[width=0.80\linewidth]{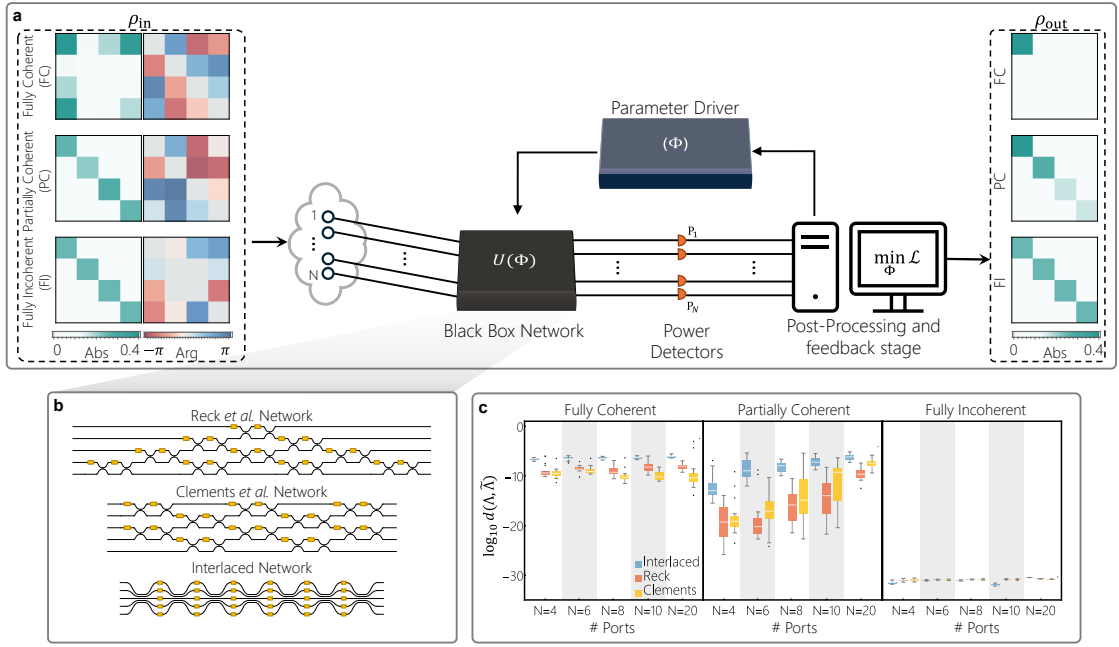}
    \caption{\textbf{Algorithmic approach and numerical performance.} \textbf{a} Schematic of the architecture-agnostic coherence matrix analyzer. A programmable unitary device (black box) is controlled by the parameter set $\Phi$. Power detectors (PD) record the output powers $P_{i}$, which are used to compute a figure of merit, $\mathcal{L}$. A parameter driver iteratively updates $\Phi$ based on these measurements until $\mathcal{L}$ is minimized. The left dashed box depicts the magnitude and phase of randomly sampled coherence matrices of various types. The right dashed box illustrates the output after the unitary network diagonalizes the input coherence matrix. \textbf{b} Example of various topologies for universal unitary PICs, comprising MZI-based networks such as Reck~\cite{reck1994experimental} and Clemments~\cite{clements2016optimal}, as well as multiport coupler networks~\cite{Markowitz23Auto,zelaya2024goldilocks,Tanomura20}. \textbf{c} Numerical simulation results showcasing the error between the original and the reconstructed eigenvalues using our architecture-agnostic approach with the universal unitary topologies shown in panel~\textbf{b}.}
    \label{fig:F1}
\end{figure*}
% -------------------------------------------

Let us assume that the partially coherent field sampled at the $N$ points is fed into a general linear transform $A\in GL(N)$. Following the definition of the coherence matrix, and assuming that the changes in time of the linear transformer are much larger than the statistical fluctuations of the fields, one can see that the coherence matrix produced at the output of the linear transformer becomes $\rho_{\textnormal{out}}=A\rho A^{\dagger}$. %Following the eigen-decomposition, one can notice that the degree of coherence remains unaffected if $A\in\mathcal{U}$. This can be used as an asset to recover the singular value of the behavior characterizing the radiation field. 
The degree of coherence is measured by the rank of the coherence matrix; however, reconstructing the full coherence matrix is challenging and, in some experimental setups, even prohibitive. In turn, it is more convenient to reconstruct only its eigenvalues, which already encode the degree of coherence and can be used to analyze the coherence content, as proven successful in Ref.~\cite{roques2024measuring}. The structure of the coherence matrix can be modified in two ways:
\begin{itemize}
\item Unitary control, $A\in U(N)$, shuffles the correlation components while preserving the degree of coherence. This is a handy resource for controlling the scattering and absorption dynamics of radiation fields before reaching the scatterer~\cite{guo2024unitaryA,guo2024unitaryB}. 
\item Introducing losses into the network modifies the eigenvalues of the coherence matrix, altering the degree of coherence. Such losses break the unitary evolution and induce a general linear transform $A\in GL(N)\supset U(N)$ instead. Still, the output density matrix transforms as $\rho_{\textnormal{out}}=A\rho_{\textnormal{in}}A^{\dagger}$ and satisfies the coherence matrix conditions.
\end{itemize}

The diagonal component $\rho_{n,n}$ is proportional to the intensities at the $n$-th sampling point, for $n\in\{1,\ldots,N\}$. The total intensity is proportional to the trace of $\rho$, which, without loss of generality, we normalize to unity ($\textnormal{tr}\rho=1$). We begin the analysis by focusing on parameterized $N$-dimensional unitary matrices, $U(\Phi):\mathbb{R}^{K}\rightarrow \mathbb{C}^{N\times N}$, which are not necessarily universal. Here, $K$ denotes the total number of real-valued parameters in $\Phi=\{\phi_{k}\}_{k=1}^{K}$ controlling the transfer matrix of the unitary network, the specifics of which depend on the network design.  

Fig.~\ref{fig:F1}\textbf{a} illustrates a typical experimental setup scenario involving a general linear transform, the input radiation field to be analyzed, and the power gathering and processing. The power measurements at the device output correspond to the diagonal components of the coherence matrix, $(\rho_{\textnormal{out}})_{nn}$. Interestingly, if the coherence matrix is already diagonal, the power measurements correspond to the singular values of the output radiation field, up to a normalization constant. From the mathematical standpoint, one can achieve such a diagonalization if the parameters $\Phi$ of the unitary processor are chosen such that $U(\Phi)=V^{\dagger}$. This reduces the output coherence matrix to $\rho_{\textnormal{out}}=\Lambda$, from which the analysis of the coherence degree becomes immediate.

The diagonalization approach has been shown to be successful in Ref.~\cite{roques2024measuring} via a photonic triangular unitary network with the topology of Reck \textit{et al.}~\cite{reck1994experimental}. The latter network is suited for this task, as its topology allows layer-wise manipulation of light, maximizing output power at the ports in a descending manner. Once the power at a given output port is maximized, the network allows for independent tuning of the subsequent layer without interfering with the previous steps. Likewise, a similar approach has been implemented in Ref.~\cite{hashemi2026chip} with a hexagonal-like photonic unit. However, complete knowledge of the network topology is not always possible, nor can its performance be ensured in non-ideal scenarios. Thus, to make the analysis of coherence matrices widely applicable across different network topologies, we propose an experimentally aware method that is independent of the unitary photonic architecture and treats the PIC as a black box.

Since the $N$ power measurements at the network output sample the diagonal components of the output coherence matrix, and the network performs unitary operations, the eigenvalues of the output coherence matrix $\rho_{\textnormal{out}}$ are preserved. Thus, to extract the coherence matrix eigenvalues, we shall ensure that the output coherence matrix is diagonal; in that case, the power measurements correspond, up to a normalization factor, to the coherence matrix eigenvalues.  Although full characterization of the coherence matrix is beyond the scope of this letter, the diagonalization condition can be assessed only by measuring the output power. This can be ensured via the \textit{Schur–Horn theorem}~\cite{horn1954doubly}: Let $X=\{ x_{1}\geq x_{2} \geq \ldots x_{N}, x_{n}\in\mathbb{R}\}$ and $Y=\{ y_{1}\geq y_{2} \geq \ldots y_{N}, y_{n}\in\mathbb{R}\}$ be two sequences of non-increasing real numbers; then, there is a Hermitian matrix $H$ with diagonal components $Y$ and eigenvalues $X$ if and only if {\small $\sum_{n=1}^{K}(x_{n}-y_{n})\geq 0$}, for all $K=\{1,\ldots,N\}$. Since coherence matrices satisfy this condition, normalized power measurements will never exceed the normalized coherence matrix eigenvalues. Furthermore, the equality in Schur-Horn's theorem holds when the coherence matrix is already diagonal.

We can leverage Schur-Horn's theorem to define an objective that converts any incident radiation field into diagonal form, without prior knowledge of the linear unitary network or the incident field. To this end, let us define the relative power vector
\begin{equation}
    \widetilde{\mathbf{P}}(\Phi)=(\widetilde{P}_{1},\ldots,\widetilde{P}_{N-1}), \quad  \widetilde{P}_{n} = 1 - \sum_{k=1}^{n}P_{k}(\Phi) , 
\end{equation}
where $P_{n}(\Phi)\equiv(\rho_{\textnormal{out}})_{n,n}$ are the normalized power measuresd at the $n$-th output port. By following the Schur-Horn theorem, the output coherence matrix $\rho_{\textnormal{out}}$ takes a diagonal form whenever the $n$-th cumulative power, $\sum_{k=1}^{n}P_{k}$, is maximal for each $n$. This is, in turn, equivalent to the minimization problem  
\begin{equation}
    \underset{\Phi\in\mathcal{S}}{\textnormal{min}} \Vert \mathbf{\widetilde{\mathbf{P}}}(\Phi) \Vert^{2} .
\end{equation}
It is worth stressing that the latter is independent of the unitary architecture topology, i.e., architecture agnostic. Such a feature is useful if the unitary network at hand has not been previously characterized or if the inner topology makes it difficult to estimate the light-travel paths in each section. For instance, in Ref.~\cite{roques2024measuring}, the topology of the triangular network is leveraged to steer the incoming radiation field in a sequential process until the diagonal matrix becomes diagonal. Likewise, the hexagonal topology in Ref.~\cite{hashemi2026chip} follows a similar reasoning. 

The simple yet robust approach presented here enables an error-resilient strategy: if individual components deviate from ideal behavior due to thermal crosstalk or other unwanted effects, our approach treats the entire device as a black box and applies compensation to the active components as needed. Furthermore, for the numerical assessment of the proposed method, the exact gradients of the objective function can be readily obtained, %\textcolor{red}{(see Supplementary Section S1)}
expediting our numerical tests, especially for systems with a large number of active parameters.

To numerically validate our architecture-agnostic approach, we consider three network topologies: Reck~\cite{reck1994experimental}, Clements~\cite{clements2016optimal}, and interlaced multiport-coupler~\cite{Markowitz23Auto,zelaya2024goldilocks,Tanomura20} architectures. For the interlaced structure, illustrated at the bottom of Fig.\ref{fig:F1}\textbf{b}, we parameterize the network by the number of layers ($M\in\mathbf{Z}^{+}$) using the transfer matrix $U_{\textnormal{int}}(\Phi)=FP(\Phi^{M})F\ldots FP(\Phi_{1})F$. $F\in U(N)$ is the unitary transfer matrix corresponding to a passive multiport coupler, which is an splits the light among its output ports without introducing gain or loss. Furthermore, $P(\Phi^{(m)})=\textnormal{diag}(e^{i\phi_{1}^{(m)}},\ldots,e^{i\phi_{N}^{(m)}})$ represents the $m$-th programmable and diagonal phase layer, where $\phi_{n}^{(m)}$ sets the controllable phase shift for the $n$-th arm of the $m$-th layer. We denote the set of phase parameters as $\Phi=\cup_{m=1}^{M}\Phi^{(m)}$.

To ensure a comprehensive evaluation across these topologies, we generate three random sets of $N$-point fully coherent (FC), partially coherent (PC), and fully incoherent (FI) matrices for various $N$, where each set contains 500 random samples. The error between the eigenvalues of the original targets ($\lambda_{n}$) and the reconstructed eigenvalues ($\widetilde{\lambda}_{n}$) is computed through the standard distance metric 

$$d(\mathbf{\Lambda},\widetilde{\mathbf{\Lambda}})=\sum_{n=1}^{N}\frac{(\lambda_{n}-\widetilde{\lambda}_{n})^{2}}{N}.$$

The statistical information of the error for each ensemble of samples is presented in the bar plots of Fig.~\ref{fig:F1}\textbf{c}. The results for the interlaced network were performed using the universal setup with $M=N+1$ layers~\cite{Markowitz23Auto}. These results show that no significant performance penalty is observed when transitioning across various network topologies and port sizes. %In a few cases, we observe outliers with higher error, which result from the optimizer getting trapped in a higher-error local minimum.   

\begin{figure}[t!]
    \centering
    \includegraphics[width=0.80\linewidth]{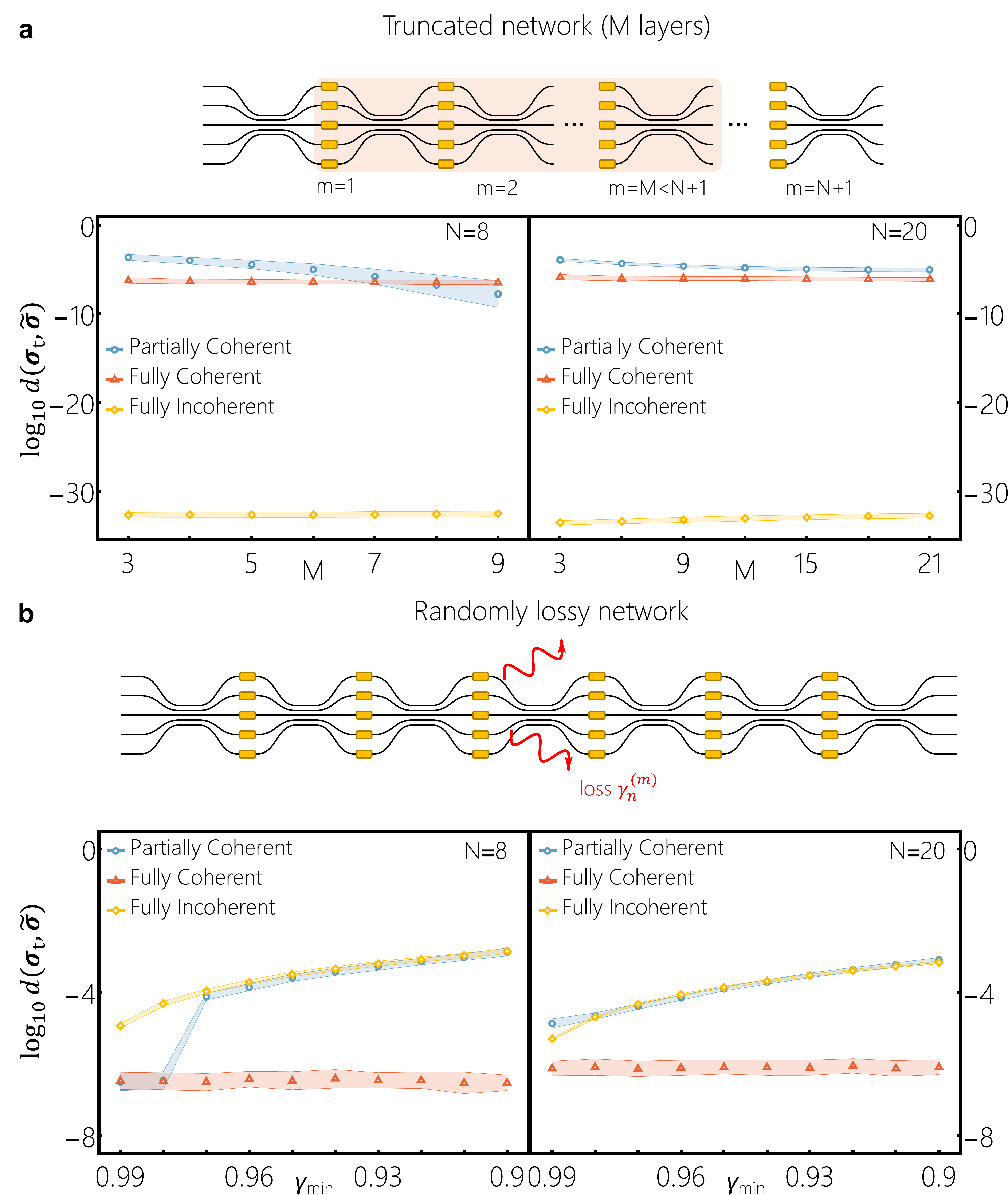}
    \caption{
    \textbf{Performance against truncation and losses.} 
    \textbf{a} Non-universal interlaced network and 
    \textbf{b} Impact of inherent arbitrary losses in the interlaced unitary network. The losses are randomly assigned in each layer by replacing the phase $\phi_{n}^{(m)} \rightarrow \phi_{n}^{(m)}-i\ln \gamma_{n}^{(m)}$, where the transmitances are randomly sampled from the interval $\gamma_{n}^{(m)}\in(\gamma_{\textnormal{min}},1)$.
    }
    \label{fig:F2}
\end{figure}

The architecture-agnostic framework of the proposed coherence analyzer proves handy when considering under-parameterized networks, since the exact diagonalization condition can be relaxed to the approximation $U(\Phi)V^{\dagger}\approx\mathbb{I}$. Indeed, in other scenarios, the universality is not required for optical computing tasks, and a lower-depth network suffices for tasks such as photonic state generation~\cite{zelaya2025chip} and matrix randomization and encryption~\cite{zelaya2025integrated}. We thus consider a truncated interlaced network with $M<N+1$ layers, as shown in Fig.~\ref{fig:F2}\textbf{a}, and perform the coherence analysis on an ensemble of 500 random samples for each combination of $N$ and $M$. We separate the analysis for ensembles of random fully coherent (FC), partially coherent (PC), and fully incoherent (FI) matrices. The objective function can extract the eigenvalue with as few as three layers for PI light, with an error of around $10^{-3}$ for $N=8$ and $10^{-4}$ for $N=20$. 
% -------------------------------------------
% -------------------------------------------
In practical applications, optical networks inevitably experience losses, particularly during routing between components and around sharp bends. While typically minimal, these losses can accumulate and degrade overall performance. However, due to its overparameterization, the interlaced network demonstrates intrinsic resilience to such imperfections and moderate losses~\cite{Markowitz2023auto}. To model this, we adapt the coherence analyzer algorithm by incorporating a loss term into each phase element $\phi_{n}^{(m)}\rightarrow\phi_{n}^{(m)}-i\log\gamma_{n}^{(m)}$, where $\gamma_{n}^{(m)}\in(0,1)$ denotes the transmission amplitude. To quantify the impact of these losses on the extraction of eigenvalues from the input coherence matrix, we optimize the network using parameters randomly sampled from $\gamma_{n}^{(m)}\in(\gamma_{\textnormal{min}},1)$. Varying $\gamma_{\textnormal{min}}$ establishes a controlled upper bound on component-level loss. For each $\gamma_{\textnormal{min}}$, we sample 200 ensembles of $N^{2}$ random loss parameters. Although real-world losses are rarely this extreme or purely random, this stochastic approach provides a worst-case assessment. The minimum bound on the transmission in each configuration is $\gamma_{\textnormal{min}}^{M}$, where $M$ is the number of layers in the interlaced architecture. Despite the losses, the output power is normalized so that the rest of the algorithm remains unmodified. Figure~\ref{fig:F2}\textbf{b} summarizes the numerical results for $N=8$ and $N=20$, utilizing the input fields (partially coherent, fully incoherent, and fully coherent) previously established in Fig.~\ref{fig:F1}\textbf{c}. The data illustrate the average error distance (markers) and standard deviation (shaded area) between the reconstructed and original eigenvalues. As anticipated, the error distance increases inversely with transmittance, scaling higher as the maximum permissible loss grows.

\textit{Conclusions} -- In this letter, we have presented an architecture-agnostic approach for the analysis of partially coherent light using programmable photonics. By relying on the Schur-Horn theorem, our method successfully extracts the eigenvalues of the coherence matrix from output power measurements alone. A significant advantage of this approach is that the device can be programmed as an identity matrix, allowing the incoming field to pass through unaffected after characterization for subsequent use. 

Beyond universal networks, we demonstrated that this black-box methodology is highly adaptable, functioning effectively even in under-parameterized architectures and in the presence of arbitrary optical losses. This versatility underscores the robustness of our approach, paving the way for space-efficient, error-resilient, and programmable spatial coherence analyzers. \\

% \subsection*{Funding} 
\noindent\textbf{Funding.} This project is supported by the U.S. Air Force Office of Scientific Research (AFOSR) Award\# FA9550-25-1-0200. \\

% \subsection*{Disclosures}
\noindent\textbf{Disclosures.} The authors declare no conflicts of interest.\\

% \subsection*{Data availability}
\noindent\textbf{Data availability.} Data underlying the results presented in this paper can be obtained from the authors upon reasonable request. 

% \subsection*{Supplemental document} 
% See Supplement Materials for supporting content.

% Bibliography
\bibliography{biblio}

\begin{thebibliography}{10}
\newcommand{\enquote}[1]{``#1''}

\bibitem{mandel1996optical}
L.~Mandel, E.~Wolf, and J.~H. Shapiro, \enquote{Optical coherence and quantum
  optics,}  (1996).

\bibitem{lee2020light}
S.~Lee, D.~Kim, S.-W. Nam, \emph{et~al.}, {\protect\JournalTitle{Scientific
  reports}} \textbf{10}, 18832 (2020).

\bibitem{reck1994experimental}
M.~Reck, A.~Zeilinger, H.~J. Bernstein, and P.~Bertani,
  {\protect\JournalTitle{Physical review letters}} \textbf{73}, 58 (1994).

\bibitem{de2018simple}
H.~de~Guise, O.~Di~Matteo, and L.~L. S{\'a}nchez-Soto,
  {\protect\JournalTitle{Physical Review A}} \textbf{97}, 022328 (2018).

\bibitem{miller2013self}
D.~A. Miller, {\protect\JournalTitle{Photonics Research}} \textbf{1}, 1 (2013).

\bibitem{yi2021multi}
D.~Yi, Y.~Wang, and H.~K. Tsang, {\protect\JournalTitle{APL Photonics}}
  \textbf{6} (2021).

\bibitem{clements2016optimal}
W.~R. Clements, P.~C. Humphreys, B.~J. Metcalf, \emph{et~al.},
  {\protect\JournalTitle{Optica}} \textbf{3}, 1460 (2016).

\bibitem{Shokraneh2020}
F.~Shokraneh, S.~Geoffroy-Gagnon, and O.~Liboiron-Ladouceur,
  {\protect\JournalTitle{Optics Express}} \textbf{28}, 23495 (2020).

\bibitem{rahbardar2023addressing}
K.~Rahbardar~Mojaver, B.~Zhao, E.~Leung, \emph{et~al.},
  {\protect\JournalTitle{Optics Express}} \textbf{31}, 23851 (2023).

\bibitem{tanomura2022scalable}
R.~Tanomura, R.~Tang, T.~Umezaki, \emph{et~al.},
  {\protect\JournalTitle{Physical Review Applied}} \textbf{17}, 024071 (2022).

\bibitem{zelaya2024goldilocks}
K.~Zelaya, M.~Markowitz, and M.-A. Miri, {\protect\JournalTitle{Scientific
  Reports}} \textbf{14}, 10950 (2024).

\bibitem{friedman2025programmable}
J.~Friedman \emph{et~al.}, {\protect\JournalTitle{Scientific Reports}}
  \textbf{15}, 35173 (2025).

\bibitem{pastor2021arbitrary}
V.~L. Pastor, J.~Lundeen, and F.~Marquardt, {\protect\JournalTitle{Optics
  Express}} \textbf{29}, 38441 (2021).

\bibitem{Markowitz2023auto}
M.~Markowitz, K.~Zelaya, and M.-A. Miri, {\protect\JournalTitle{Optics
  Express}} \textbf{31}, 37673 (2023).

\bibitem{born2013principles}
M.~Born and E.~Wolf, \emph{Principles of optics: electromagnetic theory of
  propagation, interference and diffraction of light} (Elsevier, 2013).

\bibitem{tomlins2005theory}
P.~H. Tomlins and R.~K. Wang, {\protect\JournalTitle{Journal of Physics D:
  Applied Physics}} \textbf{38}, 2519 (2005).

\bibitem{Markowitz23Auto}
M.~Markowitz, K.~Zelaya, and M.-A. Miri, {\protect\JournalTitle{Opt. Express}}
  \textbf{31}, 37673 (2023).

\bibitem{Tanomura20}
R.~Tanomura, R.~Tang, S.~Ghosh, \emph{et~al.}, {\protect\JournalTitle{Journal
  of Lightwave Technology}} \textbf{38}, 60 (2020).

\bibitem{roques2024measuring}
C.~Roques-Carmes, S.~Fan, and D.~A. Miller, {\protect\JournalTitle{Light:
  Science \& Applications}} \textbf{13}, 260 (2024).

\bibitem{guo2024unitaryA}
C.~Guo and S.~Fan, {\protect\JournalTitle{Physical Review B}} \textbf{110},
  035430 (2024).

\bibitem{guo2024unitaryB}
C.~Guo and S.~Fan, {\protect\JournalTitle{Physical Review B}} \textbf{110},
  035431 (2024).

\bibitem{hashemi2026chip}
A.~Hashemi, A.~Shiri, B.~E. Saleh, \emph{et~al.}, {\protect\JournalTitle{arXiv
  preprint arXiv:2601.18797}}  (2026).

\bibitem{horn1954doubly}
A.~Horn, {\protect\JournalTitle{American Journal of Mathematics}} \textbf{76},
  620 (1954).

\bibitem{zelaya2025chip}
K.~Zelaya, M.~Honari-Latifpour, K.~K. Mandal, \emph{et~al.},
  {\protect\JournalTitle{Optica}} \textbf{12}, 1492 (2025).

\bibitem{zelaya2025integrated}
K.~Zelaya, M.~Honari-Latifpour, and M.-A. Miri, {\protect\JournalTitle{npj
  Nanophotonics}} \textbf{2}, 6 (2025).

\end{thebibliography}

% Full bibliography added automatically for Optics Letters submissions; the following line will simply be ignored if submitting to other journals.
% Note that this extra page will not count against page length
\bibliographyfullrefs{biblio}

\end{document}